\documentclass[prd,nofootinbib]{revtex4}

\usepackage[british]{babel}
\usepackage{amsbsy}
\usepackage{amssymb}
\usepackage[sumlimits,tbtags]{amsmath}
\usepackage{graphicx}
\usepackage{subcaption}
\usepackage{xcolor}
\usepackage[active]{srcltx}
\usepackage{pdfsync}
\usepackage{xfrac}

\usepackage{mathtools}
\usepackage{ulem}

\usepackage{lipsum}

\usepackage[hyperfootnotes = false]{hyperref}
\usepackage{cleveref}
\hypersetup{
 colorlinks,
 linkcolor=blue,
 citecolor=red,
 urlcolor=blue, 
 }

\usepackage{color}

\newcommand{\ie}{i.e.,~}
\newcommand{\eg}{e.g.,~}
\usepackage{cancel}

\newcommand{\A}{{\cal A}}
\newcommand{\B}{{\cal B}}
\newcommand{\C}{{\cal C}}
\newcommand{\D}{{\cal D}}
\newcommand{\E}{{\cal E}}

\newcommand{\la}{{\langle}}
\newcommand{\ra}{{\rangle}}

\def\lna{{\hat a}}
\def\lnb{{\hat b}}

\def\lni{{\hat i}}
\def\lnj{{\hat j}}
\def\lnk{{\hat k}}
\def\lnm{{\hat m}}

\def\lnZ{{\hat 0}}
\def\lnO{{\hat 1}}
\def\lnD{{\hat 2}}
\def\lnT{{\hat 3}}

\def\O{{\cal O}}

\def\be{\begin{equation}}
\def\ee{\end{equation}}
\def\beq{\begin{equation}}
\def\eeq{\end{equation}}
\def\bea{\begin{eqnarray}}
\def\eea{\end{eqnarray}}

\def\bear{\begin{eqnarray}}
\def\eear{\end{eqnarray}}

\def\tn{{\tilde{n}}}
\def\teps{{\tilde{\varepsilon}}}
\def\tu{{\tilde{u}}}
\def\tsig{{\tilde{\sigma}}}
\def\ttheta{{\tilde{\theta}}}
\def\tvort{{\tilde{\omega}}}
\def\ta{{\tilde{a}}}
\def\tT{{\tilde{T}}}
\def\ts{{\tilde{s}}}
\def\tmu{{\tilde{\mu}}}

\def\tperp{{\tilde{\perp}}}

\def\om{{\omega}}

\def\veps{{\varepsilon}}

\def\pw{e^{ik_dx^d}}

\def\hk{\hat{k}}

\newcommand{\comment}[1]{}

\begin{document}

\title{A covariant approach to relativistic large-eddy simulations: The fibration picture}

\author{T. Celora$^1$, N. Andersson$^1$, I. Hawke$^1$ and G.L. Comer$^2$ }

\affiliation{
$^1$ Mathematical Sciences and STAG Research Centre, University of Southampton,
Southampton SO17 1BJ, United Kingdom\\
$^2$ Department of Physics, Saint Louis University,
St Louis, MO 63156-0907, USA }

\begin{abstract}
Models of turbulent flows require the resolution of a vast range of scales, from large eddies to small-scale features directly associated with dissipation. As the required resolution is not within reach of large scale numerical simulations, standard strategies involve a smoothing of the fluid dynamics, either through time averaging or spatial filtering. These strategies raise formal issues in general relativity, where the split between space and time is observer dependent. To make progress, we develop a new covariant framework for filtering/averaging based on the fibration of spacetime  associated with fluid elements and the use of Fermi coordinates to facilitate a meaningful local analysis. We derive the resolved equations of motion, demonstrating how ``effective'' dissipative terms arise because of the coarse-graining, and paying particular attention to the thermodynamical interpretation of the resolved quantities. Finally, as the smoothing of the fluid-dynamics inevitably leads to a closure problem, we propose a new closure scheme inspired by recent progress in the modelling of dissipative relativistic fluids, and crucially,  demonstrate the linear stability of the proposed model. 
\end{abstract}

\maketitle

\section{Motivation}

Fluid models inevitably involve aspects of averaging---we have to average over a large number of particles in order to describe a fluid system in terms of a small number of macroscopic quantities (in the thermodynamic sense). However, in the simplest settings we do not have to worry (too much) about the actual process of averaging. For example, the transition from particle kinetic theory to fluid model follows intuitively when the momentum distribution develops a well-defined peak. Similarly, the notion of a fluid element enters naturally on scales much larger than the individual particle mean-free paths. However, the story changes when we turn to numerical simulations and problems involving, for example, turbulence. When we consider the problem from a simulation point of view, we have to consider the scale associated with the numerical resolution. This numerical scale tends to be vast compared to (say) the size of a fluid element. For example, in the high-density core of a neutron star we would typically deal with mean-free paths of a fraction of a millimeter while the best current large-scale numerical simulations of neutron-star mergers involve a resolution of order 10 meters \cite{Kiuchi_2018}. This scale discrepancy has ``uncomfortable'' implications. In a highly dynamical situation we may not be able to resolve the full range of scales involved. Quite a lot of action can be hidden inside each computational cell. This problem is, of course, not new. It is a well-known fact that motivates the (considerable) effort going into developing large-eddy schemes in computational fluid dynamics (the subject of numerous textbooks, see for example \cite{lesbook,mcdonough,Lesieur}). 

The astrophysical significance of the problem is obvious given that many relevant situations  involve/require the modelling of hydro(-magnetic) turbulence. Again, the dynamics of neutron stars  comes to mind. A topical example is the turbulent flow caused by the Kelvin-Helmholtz instability, where the small scale dynamics drive the amplification of the magnetic field (through the dynamo effect) in binary neutron star mergers \cite{PriceRosswog,PRD.92.124034,PRL.100.191101,Zrake}. Traditionally---in the context of Newtonian theory---turbulent flows have been studied in terms of Reynolds averaged equations, which, roughly speaking, are obtained by time-averaging the fluid dynamics. This smoothing requires the modelling of  features that are not captured by the resolved flow, bringing in the need to introduce suitable ``closure conditions'' (necessary as the averaged scheme introduces more degrees of freedom than there are equations of motion). Most recent work replaces the averaging with  spatial filtering.
This strategy is often preferred because it involves less modelling (see e.g. \cite{mcdonough}), although Reynolds-type averaging is still widely used in the context of magnetic dynamos, see \cite{Brandeburg2005} for a review.

Because of the relevance for, in particular, binary neutron-star merger simulations, there have been several recent efforts to extend the ``familiar logic'' from the Newtonian setting to relativity. These range from the more formal discussion in \cite{eyink} to the actual simulations in \cite{radice1,radice2}. Also worth noting is the recent work in \cite{duez} in which the results from \cite{radice1} are contrasted with those obtained modelling the turbulent flow as effectively viscous on a larger scale (see e.g. \cite{shibata}). The general relativistic magnetohydrodynamics  merger simulations of \cite{Giacomazzo} are another relevant example. Most of these results are based on spatial filtering, with subgrid models tailored to account for  small-scale dynamics. Recently, a more refined gradient subgrid-scale model for general relativistic simulations was developed \cite{viganoNR,carrasco,viganoGR} and  applied to binary neutron-star mergers \cite{aguil}. 

In short, while there has been notable effort to carry out large-eddy simulations in relativity, the formal underpinnings for this effort are not as firmly established as one might like (the exception being the discussion in \cite{eyink}). This is the gap  we are trying to bridge here. Starting from the beginning, we bring to the fore the fundamental issues associated with any effort to ``average'' or ``filter'' in a curved spacetime.

We want to consider the problem from a covariant spacetime point of view, a key point being that the underlying principles---for both  time-averaging and space filtering--- should be the same (or at least ``similar''). Both strategies combine ``smoothing''  with suitable closure relations to determine contributions that may not be ``directly'' calculable (i.e. represented on the resolved scale).  The issues we are interested in can be approached at the level of an ``effective theory'' based on fairly simple rules, avoiding a detailed discussion of the underlying averaging/filtering process. This is a useful strategy as it leads to a relatively straightforward derivation of the dynamical equations. At the same time, one has to pay attention to the details as a number of issues come into play when we consider the problem from the covariant perspective of general relativity. In essence, we want to establish what a consistent spacetime scheme for averaging/filtering should look like, and highlight issues relating to the formulation of such a scheme. 

As we will discuss later (in \cref{sec:2parEoS}), the filtering/averaging process impacts on the interpretation of the microphysics  (i.e. the equation of state) that enter the hydrodynamic modelling, information which we would like to probe with numerical simulations. The fibration framework we are proposing is particularly motivated  by the fact that it allows for a clear connection with the underlying thermodynamics. However,
as the final aim is to develop a consistent  set-up for simulations, there are also important numerical issues to be discussed; e.g. the implicit filtering associated with numerical discretization---as opposed to the explicit filtering  that provides the focus for most of the present discussion. These will be explored in a follow-up paper \cite{fibrLESnum}, allowing us to devote our full attention to the formal/foundational issues in this first effort.

The paper is laid out as follows: in \cref{sec:AverVsFilter} we introduce the notion of time averaging and space filtering, and then use Fermi coordinates in \cref{sec:FCcoords} to formally define each procedure in a relativistic ``fibration scheme'' (based on the ``world lines'' of the ``fluid elements'' and the associated ``proper time''). In \cref{sec:average} we discuss averaging of the perfect fluid equations, and introduce an expansion for small fluctuations as this helps keep the discussion pedagogical. We then move on to  the filtering case in \cref{sec:filter}. To address the different issues step by step, in \cref{sec:average} we perform the analysis using a simple barotropic equation of state, with results for a more general model explored in \cref{sec:2parEoS}. Finally, we propose a  closure scheme and discuss the associated linear stability in \cref{sec:closure}. We provide brief conclusions in \cref{sec:remarks}. 

As we need to keep track of the relevant scales and quantities associated with different ``observers'', the notation easily gets somewhat messy. This may be inevitable, but let us try to ease the pain by explaining some of the key points from the outset.  First of all, we need to distinguish between fine-scale and coarse-scale quantities. To do so, we  use bars and angle brackets---\ie$\overline{A}$ and $\la A\ra$---for averaged and filtered quantities (respectively), obtained from the fine-scale one, $A$. However, as we will see, these quantities are not necessarily the most natural to evolve. Thus, we use tildes, \eg$\tilde A$, to identify  the evolved/resolved quantities. Finally, while discussing the linear stability of the proposed closure scheme (\cref{subsec:stability,subsec:Smagorinsky,subsec:fixSmag}), we  drop the tilde notation---as all the quantities  are then assumed to be evolved and there is no need to make the distinction---and use instead sub/superscripts to represent quantities evaluated on the background, like $A_0$. This subscript should not be confused with the spacetime indices, which are represented by latin letters $a,b,c \ldots = 0,1,2,3$ throughout.

\section{Averaging vs filtering}
\label{sec:AverVsFilter}

In order to provide the appropriate context and establish the general strategy, it is useful to briefly summarise the standard approach for  (typically incompressible) fluid dynamics in Newtonian gravity. 
Traditionally, small scale fluctuations are considered in terms of averaging, following the pioneering work of Reynolds and others \cite{lesbook}. In effect, this means that we have $A=\overline A+ \delta A$, with the fluctuations represented by $\delta A$ at each spacetime point. Introducing this formal split has the advantage of providing a straightforward derivation of the dynamical equations and a relatively clear interpretation of the involved quantities. One may also resort to an expansion for ``small'' $\delta A$ (see \cite{duez} for a relevant example of this). Typically, progress is made by assuming that the average of the linear fluctuations vanishes, which may not be a faithful representation of the physics the model aims to describe (see \cite{mcdonough} for a more extensive discussion). Noting this, the typical strategy for spatial filtering---forming the basis for modern large-eddy simulations---is different. In particular, the filtered fluctuations are not taken to vanish, nor does the argument involve expanding in the fluctuations. Instead, one typically proceeds by introducing a new set of variables to represent the filtered dynamics. From the conceptual point of view, each of the two strategies has attractive features and---as we are interested in the formal aspects of the problem---we will consider both of them in the following.

Let us first consider the standard averaging problem.
Without going into detail we can derive the averaged equations by applying a simple set of rules. In essence, we assume that (using a bar over quantities to represent averaging) 
\begin{subequations}\label{eq:OperationalRules}
\begin{align}
    \overline c &= c \ , && \mbox{for\ constants} \ , \\
    \overline{ A + B} &= \overline A  + \overline B \ , && \mbox{linearity of the procedure} \ ,   \\
    \overline{ \partial_a A}  &= \partial_a \overline A \ , && \mbox{averaging commutes with derivatives} \ .
\end{align}
\label{assum0}
\end{subequations}
It is immediately clear that, while the last of these relations is intuitive for a time average in Newtonian physics, we need to tread carefully when we turn to the relativistic setting. First of all, we need to face the fact that we do not have an observer-independent space-time split. Secondly, the derivatives we need to consider will be covariant, and hence must consider the spacetime curvature. Simply noting these reservations for the moment (they will be discussed in \cref{sec:FCcoords}), the stated rules imply that
\be 
\overline{ cA}  = c\, \overline A \ .
\ee
Moreover---and this is where the main distinction from large-eddy models comes in---it is common to further assume that the average of the fluctuations vanishes so we have
\be\label{eq:AverFluct}
\overline{ \delta A} = 0 \ .
\ee
It then follows that
\begin{equation} \label{eq:DoubleAverage}
    \overline{\overline{A}} = \overline{A} \;,
\end{equation}
which means that the field $\overline{A}$ remains unchanged after the averaging.
In effect, this additional rule leads to 
\begin{equation}
    \overline{ \overline A B} = \overline A\, \overline B  \;.
    \label{five}
\end{equation}
This simplifies the discussion considerably as we can ignore all linear fluctuation terms in the averaged equations. 

Time averaging is the (conceptually) simplest approach to the problem, but (strictly speaking) it removes dynamical features associated with the fluctuations\footnote{The averaged quantities in \eqref{eq:DoubleAverage} and \eqref{five} are strictly speaking not time dependent. This is easy to demonstrate for simple averaging procedures. Nevertheless, the time derivative term is usually retained in the evolution of Reynolds-Averaged Navier-Stokes equations (see \cite{mcdonough} and the comments below in \cref{sec:FCcoords}). It is worth noting this formal inconsistency, even though it does not affect the model we  develop here. }, which is unlikely to be realistic. A faithful representation of the physics may require a different prescription. One option would be to not introduce the assumption from \cref{eq:DoubleAverage}.  The typical  description then involves (spatial) filtering, using some specified kernel to define the separation of scales \cite{lesbook}. This (effectively) leads to the same kind of rules as before---with the exception of \cref{eq:AverFluct,eq:DoubleAverage,five}---although we now have (indicating filtering by angle brackets) 
\begin{equation}
    \langle \langle A\rangle B \rangle \neq \langle A\rangle \langle B \rangle \ ,
\end{equation}
which means that filtered fluctuations do not have to vanish. That is, in general we have $\langle \delta A \rangle \neq 0$. 

\section{The spacetime view: Fermi Coordinates}
\label{sec:FCcoords}

As a first---and essential---step towards a relativistic model for averaging/filtering, we have to consider the spacetime aspects of the problem. In particular, we need to introduce an unambiguous space-time decomposition---otherwise we cannot meaningfully consider ``time'' averages or ``space'' filtering. This is more than semantics \cite{eyink}. An interesting discussion of the problem (mainly from the cosmology perspective) has been provided by Ellis \cite{EllisInhomCosmo}, and it is evident that the issue is conceptually problematic since the notions of time and space are observer dependent. The problem is particularly vexing for a foliation based approach to spacetime  (as assumed in numerical relativity \cite{duez}, where the spacetime foliation is manifestly gauge dependent). However, for a fluid there does exist a natural fibration of spacetime \cite{livrev}. If we take the associated fluid frame as our starting point, we can introduce a meaningful ``local analysis'' which allows us to make progress. Moreover, it is natural to use the fluid frame to make the all-important connection with the  microphysics and the equation of state \cite{livrev}. The strategy also allows us to consider thermodynamical aspects of the averaging/filtering scheme. 

Let us explore the steps involved in an averaging/filtering procedure based on a spacetime fibration. In particular, we want to establish under which conditions we may assume that the covariant derivative commutes with the averaging/filtering procedure. Intuitively, we need to assume from the outset that there is a separation of scales between the metric fluctuations and the fluid fluctuations. The natural approach to the problem involves Fermi-type coordinates. In order to establish the logic, consider the following situation: The fluid four-velocity (and other physical properties) varies over a resolved spacetime region (this can be thought of as a numerical cell, even though such numerical cells would typically be defined in terms of a foliation). However, we assume that it is still possible to identify a family of observers associated with  a four-velocity vector field $U^a$ which can be taken to be constant over the resolved region and which  is ``close enough'' to the actual fluid four-velocity. (The latter assumption is not strictly required for the definition of an averaging procedure, nor for the spacetime decomposition, but it helps develop the logic). Then we can use the worldlines with tangent $U^a$ to construct  Fermi-type coordinates and explore the details of a given averaging/filtering procedure. 

Fermi coordinates were first introduced by Fermi in $1922$ \cite{fermi1,fermi2} and then developed by, in particular, Manasse and Misner \cite{manasse} (see also, for example, \cite{synge,rakh}). We will not dwell on the construction itself here. Instead, we focus on the properties and  region of validity of the associated coordinate system. The set of coordinates is essentially built from a spacetime tetrad transported along a central worldline (naturally taken to be timelike in our case). This is convenient because the metric and the Christoffel symbols take a very simple form along the central curve. 

Let us introduce Fermi coordinates $x^\lna = \{x^\lnZ,\,x^\lnO,\,x^\lnD,\,x^\lnT\}$ (distinguished by hats on the indices) such that, on the central worldline $G$, the metric reduces to the Minkowski form $g_{\lna\lnb}= \eta_{\lna\lnb}$ while its first derivatives can be obtained from the  Christoffel symbols (see \cite{GravitationMTW}) 
\begin{subequations}
\begin{align}
	g_{\lna\lnb,\lnZ} &=   0 \;, \\
    g_{\lnZ\lnZ,\lnj} &=  -2 a_\lnj \;,\\
    g_{\lnZ\lnj,\lnk} &= 0\;,\\
    g_{\lnj\lnk,\lnm} &=  0 \;,
\end{align}	
\end{subequations}
where the commas represent partial derivatives.
We have introduced the non-vanishing piece of the four acceleration, $a_\lnj$, of the worldline\footnote{That is, the four acceleration is $a_b = U^a\nabla_aU_b$ here.} and chosen to construct the tetrad in such a way that the associated observer  is non-rotating (which seems natural). With this construction we can formulate an expansion of the metric in the neighbourhood of the worldline. This leads to 
\begin{subequations}
\begin{align}
	g_{\lnZ\lnZ} &= g_{\lnZ\lnZ}\big|_G + g_{\lnZ\lnZ,\lna} x^\lna = -(1 + 2a_\lnj x^\lnj) + \mathcal{O}(x^\lnj)^2 \;,\\
    g_{\lnZ\lnj} &= g_{\lnZ\lnj}\big|_G + g_{\lnZ\lnj,\lna} x^\lna =  \mathcal{O}(x^\lnj)^2 \;,\\
    g_{\lni\lnj} &=   g_{\lni\lnj}\big|_G + g_{\lni\lnj,\lna} x^\lna = \eta_{\lni\lnj}  + \mathcal{O}(x^\lnj)^2 \;,
\end{align}	
\end{subequations}
where $|_G$ indicates that the quantity is evaluated on the worldline.
This is just a Taylor expansion for the metric where the ``small parameter'', $\lambda$ (say), is taken to be the proper distance from the central curve. That is, we have $\lambda^2 = (x^\lnO )^2 + (x^\lnD )^2 + (x^\lnT )^2$. We see that, if the worldline is a geodesic then $a^\lnj = 0$ and there are no corrections up to second order in the metric. However,  there will always be corrections at second order due to the spacetime curvature. These corrections can be expressed in terms of the Riemann tensor (again evaluated on the worldline $G$), but we will not need the explicit results here. Because we are assuming that the metric fluctuations happen on a larger scale (with respect to the fluid variations), we can make use of these expansions in the following.

Next, we can use the coordinates we have introduced to define a formal averaging/filtering procedure. 
Focusing on time-averaging first, we may use the spacetime split associated with the coordinates and define the procedure as 
\begin{equation}
	\overline A (\hat x) = \lim_{T\to\infty}\frac{1}{T}\int_{0}^{ T}  d\hat \tau A(\hat x,\hat \tau)  \;.
\end{equation}
That is,  given a point on the fluid element trajectory  we average in the proper time associated with the worldline. In terms of the Fermi coordinates, the time coordinate is exactly the proper time of the central curve. Note that there is no problem in taking the limit $T \to \infty$ since the Fermi coordinates are formally defined over the entire worldline. The region of validity is only limited in the spatial directions orthogonal to the central curve. From the definition, it is clear that time-averaged quantities must be time-independent, and it immediately follows that $\overline{\overline A}= \overline{A}$ (making contact with \eqref{eq:DoubleAverage}). It also implies that we should (strictly) neglect time derivatives in the averaged equations. The upshot is that the time-averaging strategy is (formally) valid for stationary flows only. The same is true in the Newtonian context---even though time derivatives are typically retained in the equations\footnote{See \cite{mcdonough} for a more extensive discussion on this.}. In fact, this is one of the main motivations in favour of spatial filtering and large-eddy models. In the following, we follow the ``convention'' and retain terms involving time derivatives, as the main point of our discussion of the time-averaging case is pedagogical. 

As an application, let us consider the averaged metric. From the metric expansion above, we immediately see that the ``time-time'' component gains a correction due to the acceleration, which does depend on the proper time. However, we only need to integrate over points on the central worldline, where we have the Minkowski metric (in Cartesian coordinates) by construction. The situation is similar for all the remaining components. 

As a result, each component of the averaged metric takes the non-averaged value from the central worldline. That is, the metric is constant (in the sense described in \cref{sec:AverVsFilter}), and we have

\begin{equation}
\overline g_{\lna\lnb} = g_{\lna\lnb} \;.
\end{equation}
Similarly, $U^a$ is constant under averaging. To see this it is sufficient to note that (in terms of the Fermi coordinates) we have $U^\lna = (1,0,0,0)^\top$ so that $\overline U^a = U^a$. 

Analogously, we can use the spacetime split to define a space-filtering. We first have to assume that the width $L$ of the region over which we are filtering---the ``resolved box''---is small enough (in terms of the distance $\lambda$)  that the Fermi coordinates are well defined on it. For instance, one such condition is $L < 1/a$ where $a$ is the magnitude of $a^\lnj$ (see \cite{Nesterov} for a detailed discussion). In this case, the  filtering procedure may be defined through 
\begin{equation}
	\la A\ra (\hat x, \hat t) = \int dV A(\hat x+\hat y,\hat t)f(\hat y)   \ , 
\end{equation}
where we introduced the filter $f$ (normalised over the resolved box). The expression simplifies further if we note that, by construction,  the spacetime point $(\hat x,\hat t)$ lies on the central worldline $(x^\lnZ =\hat t = \tau,\, x^\lni =0)$ where $\tau$ is the relevant proper time. Also, in terms of the Fermi coordinates, the volume element is
\begin{equation}
	dV = U^\lnZ \sqrt{-g} dy^\lnO  dy^\lnD  dy^\lnT = (1 + 2a_\lnj y^\lnj)^{1/2} dy^\lnO  dy^\lnD  dy^\lnT \approx (1 + a_\lnj y^\lnj) d^3\hat y \;.
\end{equation}
Note that we have not specified the exact filter to be used. This is not required at this stage, but we will assume the filter to be an even function (noting that this is the case for the three most common filters used in large-eddy models, see \cite{lesbook} for instance) and normalized over the resolved box such that we have\footnote{If the filter has a sharp boundary---i.e. vanishes at the boundary of the resolved box---the argument does not involve extending the spatial integral to infinity. However, one can think of filters with no sharp boundary, like a Gaussian filter (see \cite{mcdonough}), which may give rise to formal issues. In practice though, the exponential tail of the Gaussian should suppress anything beyond the Fermi coordinate boundary.}
\begin{equation}
	\int d^3\hat y \,f(\hat y) = 1 \quad\,,\quad\int d^3\hat y \,y^\lnj f(\hat y) = 0 \ .
\end{equation}

Again, let us first apply this  to the filtered metric. Since there are no first-order corrections in the  expansion, $g_{\lnZ\lni}$ and $g_{\lni\lnj}$ are constant over the box, and we have 
\begin{subequations}
\begin{align}
\la g_{\lnZ\lni} \ra & = 0 =  g_{\lnZ\lni} \big|_G \ , \\
\la g_{\lni\lnj}\ra &= \int d^3\hat y(1 + a_\lnj y^\lnj)  f(\hat y)  \eta_{\lni\lnj} = \eta_{\lni\lnj} =g_{\lni\lnj} \big|_G \ .
\end{align}
\end{subequations}
We also have
\begin{equation}
\begin{split}
	\la g_{\lnZ\lnZ} \ra &= -\int d^3\hat y(1 + 2a_\lni y^\lni)^{3/2}f(\hat y)  \\
    &= -1 - 3a_\lni \int d^3\hat y\,y^\lni f(\hat y)  = -1 = g_{\lnZ\lnZ}\big|_G \; \ ,
\end{split}
\end{equation}
where the last integral vanishes because of the assumed symmetry of the kernel. Once again, each component of the filtered metric takes the  non-averaged value from the central worldline throughout the region under consideration. That is, the metric is constant
\begin{equation}
	\la g_{\lna\lnb}\ra =  g_{\lna\lnb} .
\end{equation}
We also note that, by construction $U^a$ is constant over the box so we have $\la U^a\ra = U^a$.

Finally, since we have shown that the metric can (effectively) be taken to be constant under both averaging and filtering, it is easy to show that partial derivatives commute with each procedure. That is (obviously connecting with \eqref{assum0})
\begin{equation}\label{eq:partialscommute}
	\partial_a \la A\ra = \la\partial_a A \ra \quad \text{and} \quad \partial_a \overline A = \overline {\partial_a A} .
\end{equation}
We now show that these relations hold given the definitions of the average/filter above.
Let us start with the time-averaging case. When we consider the partial derivative with respect to the spatial coordinates the argument is straightforward, as the partial derivative (in the spatial direction) can be brought inside the integral. For the time derivative, we have 
\begin{equation}
	\overline{\partial_{\hat t} A}(\hat x,\hat t) = \lim_{T\to\infty} \int_0^T d\hat\tau \partial_{\hat\tau} A(\hat x,\hat\tau) =  \lim_{T\to\infty} \frac{A(T) - A(0)}{T} = 0 \;.
\end{equation}
On the other hand, if we take the time derivative of the averaged quantity, this trivially vanishes as it is time-independent. 

The argument for the time derivative in the filtering case is similarly straightforward. For spatial derivatives, we have on the one hand
\begin{equation}
	\la \partial_{\hat i} A\ra (\hat x,\hat t) = \int dV \left(\partial_{\hat i} A\right)(\hat x+\hat y,\hat t)f(\hat y) \;,
\end{equation}
while, on the other hand,
\begin{equation}
	\frac{\partial}{\partial x^{\hat i}} \la A\ra (\hat x,\hat t) \big|_{\hat x_0}= \int dV  \frac{\partial}{\partial x^{\hat i}} A(\hat x+\hat y,\hat t)\big|_{\hat x_0} f(\hat y) \;.
\end{equation}
Using the  chain-rule in the last equation, we see that
\begin{equation}
	\frac{\partial}{\partial x^{\hat i}}A(\hat x+\hat y)\big|_{\hat x_0} = \frac{\partial A(\hat z)}{\partial z^{\hat i}}\big|_{\hat z = \hat x_0+ \hat y} \ , 
\end{equation}
and it is clear that the two relations lead to the same result.

We  now have all the ingredients we need to prove that covariant derivatives commute with the averaging/filtering procedure. In fact, given that the metric is constant (in the sense of \cref{eq:OperationalRules}), we have 
\begin{subequations}\label{eq:gabprimeconst}
\begin{align}
		\la \partial_c g_{ab}\ra &= 	\partial_c  \la g_{ab}\ra = \partial_c g_{ab} \;, \\
        \overline{\partial_c g_{ab}} & = \partial_c \overline{g_{ab}} = \partial_c g_{ab} \;.
\end{align}
\end{subequations}
As a result, the Christoffel symbols---which are obtained from combinations of first derivatives of the metric---are (locally) constant under the procedure, as well. Therefore, we have 
\begin{equation}
	\la\nabla_a A^b \ra = \partial_a \la A^b \ra + \Gamma^b_{ac} \la A^c\ra = \nabla_a \la A^b\ra \,,
\end{equation}
with an analogous result for the time-averaging case. At the end of the day, the argument is quite intuitive.

\subsection{On covariance and the Einstein equations}

It makes sense now, before we move on, to spell out the covariance of the proposed averaging/filtering procedure, and comment on its compatibility with the field equations of general relativity. It is, in fact, clear that the integrals we used  to define the averaging/filtering procedure are not of the usual type. We have to define each procedure in such a way that the   integration preserves the tensorial nature of the input\footnote{Recall that the usual integral on a (sub-)manifold of dimension $p$ takes a $p$-form as input and outputs a scalar.}. 
Given this, we define the procedure for scalar quantities and then apply it to each component\footnote{Intuitively, because the procedures are based on the Fermi-coordinates construction in terms of a non-rotating tetrad with respect to $U^a$, we can effectively think of the components as scalars.} of, say, the metric tensor. We then require the averaged/filtered quantity to transform as a tensor on the resolved scale. The proposed definition reduces to the one of \cite{eyink} in the special relativity context and it also leads back to the Newtonian ones  \cite{lesbook,mcdonough}. The main difference is that, in special relativity such integrals are chart independent---as long as the integration is performed on each component using a fixed basis (see \cite{GourghoulonSR})---while this is not the case in general relativity. 
However, even though the special relativistic integrals
are chart independent, the results are not, because the notions of length- and time scales are observer dependent. We also note that, as shown in \cite{eyink}, the observer-dependence cannot be resolved by the introduction of some kind of ``spacetime'' filter. However,  we are setting up the averaging/filtering using Fermi coordinates defined from the fibration. As this is
naturally associated with the fluid motion, the ``gauge'' dependence of the procedure is more physical. We execute the smoothing in ``some'' local frame $U^a$ which we can choose to ``associate'' to the ``micro-scale'' fluid motion.

Let us now discuss the averaging/filtering of the Einstein equations, focusing on the geometry. First of all, consider the Einstein tensor $G^{ab}$. From \cref{eq:partialscommute,eq:gabprimeconst} it follows that
\begin{subequations}
\begin{align}
		\la  g_{ab,cd}\ra &=  g_{ab,cd} \;, \\
        \overline{ g_{ab,cd}} &=  g_{ab,cd} \;.
\end{align}
\end{subequations}
Since the Einstein tensor is ultimately a combination of the metric and its (up-to-second order) derivatives $\boldsymbol{G}=\boldsymbol{G}(g,\,\partial g,\, \partial^2 g)$, this implies that we must have
\begin{equation}
    \overline{\boldsymbol{G}(g,\,\partial g,\, \partial^2 g)}  =\boldsymbol{G}(\overline g,\,\partial \overline g,\, \partial^2 \overline g) = \boldsymbol{G}(g,\,\partial g,\, \partial^2 g) \;,
\end{equation}
and analogously for the filtering case. The net result is that the coarse-grained theory remains consistent with general relativity. In particular, the Einstein equations become
\begin{subequations}
\begin{align}
    G^{ab} &= 8\pi \overline{T^{ab}} \;, \\
    G^{ab} &= 8\pi \la T^{ab}\ra \;.
\end{align}
\end{subequations}
These results follow from the Fermi-coordinate construction, and the assumed separation of scales in the metric fluctuations with respect to the fluid variables. This should be a safe assumption for binary neutron star merger applications, but not necessarily for problems relating to the  very early universe (where quantum fluctuations in the gravitational field may play an important role). In this sense the Fermi-coordinate construction should be regarded as a pragmatic argument rather than a mathematical proof.

We can now focus the discussion on the matter side, i.e. the stress-energy-momentum tensor $T^{ab}$. It is also worth noting the  obvious implication $\la T^{a b} \ra = \overline{T^{a b}}$, although it is not clear to what extent this observation is relevant in practice given that the identified observers ($U^a$) are unlikely to be the same in the two cases.

\section{Averaging in the fluid frame}
\label{sec:average}

Backed up by the Fermi-coordinate argument, let us first explore the problem of spacetime averaging. We start by introducing a fine-grained congruence  of wordlines with tangent vector field $u^a$, e.g. associated with individual fluid elements. Then,  working on a slightly larger (coarse grained) scale, we introduce another vector field $U^a$---the one used to define Fermi coordinates---such that small scale features are smoothed.  We can then use the decomposition
\begin{equation}
u^a = \gamma (U^a + \delta v^a) \ , 
\end{equation}
with
\begin{equation}
    U_a \delta v^a = 0  \ ,
  \label{con0}
\end{equation}
and
\begin{equation}
\gamma = \left( 1 - \delta v^2 \right)^{-1/2} \approx 1 + \frac{1}{2} g_{ab} \delta v^a \delta v^b \ .
\end{equation}

This coarse grained field $U^a$ is obviously not uniquely defined; any choice such that the fluctuations $\delta v^a$ can be considered ``small'' is acceptable.
In fact, at this point we take the view that it is natural to assume $\delta v\ll 1$ (the speed of the fluctuations is well below that of light). This should be a safe assumption for the problems we are interested in \cite{duez}. This allows us to develop the logic more explicitly, even though we will drop this assumption later. One may  view the linear assumption as an additional  constraint (alongside the assumptions of the Fermi frame) on the size of the region we average over, although we will not try to make this statement precise. Finally, let us assume that it makes sense to work with an ordered expansion in the fluctuations. Working to second order---throughout the discussion of averaging but not in the filtering case that follows, where the expressions are not expanded in this sense---we then have
 \begin{equation}
u^a \approx  \left( 1 + \frac{1}{2} g_{bc} \delta v^b \delta v^c\right) U^a + \delta v^a \;.
\label{bigU}
\end{equation}

Let us now consider the average of this four velocity. Given the set-up we have  $\overline U^a = U^a$ and one might expect to have $\overline u^a = U^a$, as well. However, the problem turns out to be a little bit more intricate than that. First of all, from the discussion of time-averaging in section \ref{sec:AverVsFilter} we assume 
\begin{equation}\label{eq:bardeltava}
    \overline{ \delta v^a } = 0  \ .
\end{equation}
It is also worth noting that the averaging procedure preserves directionality; that is 
\begin{equation}
    U_a \delta v^a = 0 \ \Longrightarrow U_a  \overline{\delta v^a} = 0 \;.
\end{equation}
This holds as long as we satisfy the conditions laid out in \cref{sec:FCcoords}---not because of \cref{eq:bardeltava}. 
We also get (to second order)
\begin{equation}
    \overline \gamma = 1 + \frac{1}{2} g_{ab} \overline{ \delta v^a \delta v^b } \ ,
\end{equation}
and it follows that
\begin{equation}
\overline {u}^a = \left( 1 +  \frac{1}{2} g_{bc} \overline { \delta v^b \delta v^c} \right) U^a
= \overline \gamma U^a \ .
\end{equation}

At this point we reach an impasse. It is clear that $\overline u^a$ is not (automatically) normalised and therefore can not serve as a four velocity. We would have to re-calibrate co-moving clocks to depend on the averaged fluctuations. This is problematic as we need  a projection to effect the space-time split and one would expect this to involve the fluid four velocity. There seems to be two ways to proceed. First, we could (perhaps pragmatically) opt to work with $U^a$ as the variable representing the flow. Alternatively, we may constrain the fluctuations to ensure that the averaging procedure returns $\overline u^a=U^a$. This would follow if we were to assume  the fluctuations to be such that 
\begin{equation}
   g_{ab} \overline { \delta v^a \delta v^b} = 0 \ \Longrightarrow \ \overline \gamma =1 \ .
   \label{con1}
\end{equation}
This allows us to move on, working with $\overline u^a$ to represent the flow, which  might seem the natural generalisation of the  Newtonian logic. However, considering the expected nature of small-scale turbulence, condition \eqref{con1} seems too restrictive. By assuming that the variance of the velocity fluctuations vanishes, we effectively remove the small scale kinetic energy that links to large-scale features in standard eddy-based models for turbulence \cite{lesbook,mcdonough}. 
The second of the suggested approaches thus seems unattractive and we will not pursue it further. 
A third---indeed, likely preferred---possibility becomes apparent when we consider the equation for the conserved matter flux.

\subsection{Baryon number conservation}\label{subsec:BaryonConserv_aver}

Having discussed the issues associated with averaging the four-velocity, the natural next step is to consider baryon number conservation. Letting $n = \overline n + \delta n$ represent the baryon number, the matter flux takes the form
\begin{equation}
    n^a = n u^a \approx \left( \overline n + \delta n \right)  ( \gamma  U^a + \delta v^a) \ , 
\end{equation}
such that (since $\overline{ \delta n} =0$ from the averaging) 
\begin{equation}
    \overline n^a = \overline n\, \overline \gamma \, U^a + \overline{ \delta n \delta v^a} \ .
\end{equation}
The number density measured by an observer moving along with $U^a$ is then given by
\begin{equation}
     n_0 = - U_a \overline n^a = \overline n\, \overline \gamma \ , 
\end{equation}
and we can write the averaged flux as
\begin{equation}
    \overline n^a = n_0 U^a + \overline{ \delta n \delta v^a} \ .
\end{equation}

The continuity equation then becomes
\begin{equation}
 \overline{\nabla_a n^a} = \nabla_a \overline n^a = \nabla_a (  n_0  U^a) + \nabla_ a \left( \overline{ \delta n \delta v^a }\right) = 0 \ ,
\end{equation}
or 
\begin{equation}
U^a\nabla_a n_0 + n_0 \nabla_a  U^a = - \nabla_ a \overline{ \delta n \delta v^a } \;.
 \label{bary1}
\end{equation}
This last equation shows that there is particle diffusion at second order (relative to $U^a$). Fluctuations lead to drift from large scale elements to their neighbours.

While nothing prevents us from taking this as given and moving on to the stress-energy-momentum tensor, the other equations of motion and the equation of state, it is useful to consider how the particle diffusion depends on our choice of observer $U^a$ by considering a density-weighted velocity. The advantage of this is that we can arrange things in such a way that the new four velocity is normalised, while the non-linear fluctuations are hidden in its definition. In essence, we use the weighting with the number density to adjust the co-moving clocks in the desired way. Suppose we define 
\begin{equation}\label{eq:tilde_na}
    \overline n^a = \tilde n \tilde u^a \ , 
\end{equation}
while insisting that $\tilde u_a \tilde u^a = -1$. This immediately leads to\footnote{From \cref{eq:tilde_na,eq:tilde_n} we see that this density-weighted average corresponds to the Favre-type averaging often used in the Newtonian context (see, for instance, \cite{Lesieur,SchmidtLES,dartevelle2005comprehensive}).}
\begin{equation}\label{eq:tilde_n}
    \tilde n = \overline n \,  \overline \gamma \ , 
\end{equation}
and
\begin{equation}
    \tilde u^a = U^a + \frac{1}{\tilde n} \overline {\delta n \delta v^a}
    \label{FavreU} \ .
\end{equation}
It is easy to see that this retains the required normalisation (as long as we ignore terms beyond second order). Crucially, this would remain true (by construction) if we did not expand in small fluctuations. 

Before we move on, it is important to note a difference from the Newtonian case. The new variable $\tilde u^a$ in \eqref{eq:tilde_na} becomes uniquely defined only after the introduction of $U^a$. This is evident from \eqref{FavreU} where the second term on the right-hand side explicitly depends on the time-average, which in turn relies on\footnote{The argument may seem to rest (perhaps uncomfortably so) on the somewhat arbitrary nature of $U^a$, which could be seen as an undesirable feature of the scheme. However, in practice one would not build the model this way (we are being explicit for pedagogical reasons). Pragmatically, the weighted observed $\tilde u^a$ can be simply defined by the absence of closure terms in the equation for baryon number conservation, see equation \eqref{bary2}.} the space-time split associated with $U^a$. 

We can now meaningfully introduce the projection
\begin{equation}
    \tilde \perp^a_{\,b} = g^a_{\,b} + \tilde u^a \tilde u_b \ .
\end{equation}
As anticipated, it also follows that 
\begin{equation}
    \tilde n = - \tilde u_a \overline n^a \ , 
\end{equation}
and the continuity equation takes the form
\begin{equation}\label{bary2}
    \nabla_a ( \tilde n \tilde u^a ) = \dot{\tilde n}  + \tilde n \nabla_a \tilde u^a = 0 \ ,
\end{equation} 
where (here and in the following) the ``dot'' represent the covariant derivative with respect to $\tu^a$, i.e. $\dot{\tilde n} = \tu^a\nabla_a\tn$. This is attractive because---as the fluctuations are ``hidden''---we are left with a conservation law of the pre-averaged form. However, this step comes with an important caveat.  We can remove the need for a closure in this equation, but this forces us to use the $\tu^a$ observer. One can imagine preferring a different choices of observer, which would then lead to drift terms entering  the continuity equation, as in \eqref{bary1}.

\subsection{Averaged matter dynamics}

Next, consider the perfect fluid stress-energy-momentum tensor. Starting from 
\begin{equation}
T^{ab} = (p+\varepsilon) u^a u^b + p g^{ab} \ , 
\end{equation}
we use the expansion of $u^a$ in terms of $U^a$ and the fluctuations $\delta v^a$ in \cref{bigU} (noting that, to second order in the fluctuations, we have $\overline{\gamma^2} = \overline \gamma^2$), then after averaging we introduce $\tilde u^a$ according to the density weighted Favre-average in \eqref{FavreU}. This leads to
\begin{equation}
\overline T^{ab} = 
(\overline p+ \overline \varepsilon )\overline \gamma^2 \tilde u^a \tilde u^b  + \overline p g^{ab} + 2 \tilde u^{(a} q^{b)} + s^{ab} \ , 
\end{equation}
with 
\begin{equation}\label{eq:average_q}
    q^a = -\frac{\overline p + \overline \varepsilon}{\overline n} \overline {\delta n \delta v^a} + \overline{ (\delta p + \delta\varepsilon) \delta v^a} \;,
\end{equation}
and
\begin{equation}
    s^{ab} = (\overline p + \overline \varepsilon) \overline{\delta v^a \delta v^b} \;.
\end{equation}
It is convenient to work with the energy density measured by an observer moving along with $\tilde u^a$. This follows from 
\begin{equation}
    \tilde \varepsilon = \tilde u_a \tilde u_b \overline T^{ab} =  \overline \gamma^2\overline \varepsilon + \left( \overline \gamma^2 - 1\right) \overline p \;.
\end{equation}
As a result, we have
\begin{equation}
    \overline p + \tilde \varepsilon = ( \overline p + \overline \varepsilon) \overline \gamma^2 \;,
    \label{newp}
\end{equation}
which means that we can rewrite the stress-energy-momentum tensor as
\begin{equation}
\overline T^{ab} = 
(\overline p+ \tilde \varepsilon )\tilde u^a \tilde u^b  + \overline p g^{ab} + 2 \tilde u^{(a} q^{b)} + s^{ab} \;.
\label{tabav}
\end{equation}
The equations of motion 
\begin{equation}
\nabla_a \overline T^{ab} = 0 \;,
\end{equation}
then lead to the energy equation;
\begin{equation}
\dot\teps + (\overline p + \tilde \varepsilon) \nabla_a \tilde u^a = \tilde u_b \tilde u^a \nabla_a q^b - \nabla_a q^a + \tilde u_b \nabla_a s^{ab} \;,
\label{enav}
\end{equation}
and the momentum equation;
\begin{equation}
    (\overline p + \tilde \varepsilon)\,\ta^b + \tilde \perp^{ab} \nabla_a \overline p = -  \tilde \perp^b_{\ c} \tilde u^a \nabla_a q^c - q^b \nabla_a \tilde u^a - q^a \nabla_a \tilde u^b - \tilde \perp^b_{\ c} \nabla_a s^{ac}\;,
\label{momav}
\end{equation}
where $\ta^b = \tu^a\nabla_a\tu^b$.

It is worth remarking that with this choice of resolved variables, the final equations of motion reminds us of a general dissipative fluid with viscosity and heat-flux~\cite{livrev}. The upshot is that we have to deal with dissipative effects at some level. Moreover, the  fact that we end up with an effective stress-energy-momentum tensor, highlights the intimate relation between filtering and  coarse-graining leading to an effective field theory. However, as detailed in~\cite{KovtunStable}, there are ambiguities in the construction of non-ideal relativistic fluids, and we need to fix these (through a choice of frame, in the terminology of~\cite{KovtunStable}), to make the interpretation of this model as a non-ideal fluid more precise. The first ambiguity is the meaning of the coarse-grained four velocity, which we have fixed through the use of the weighted average in~\eqref{FavreU}. Other ambiguities are linked to the out-of-equilibrium meaning of temperature and entropy currents. To fix these, we need to consider our choice of coarse-grained equation of state. This step inevitably involves additional assumptions. We will see that in simple cases (see \cref{subsec:BarotropicEoS}) the same functional form as the fine-grained equation of state may suffice, but in the general case (see \cref{sec:2parEoS}) further assumptions about the equation of state and its links to the effective entropy and temperature are needed in order to precisely interpret the coarse-grained model as a non-ideal fluid.

\subsection{The equation of state}\label{subsec:BarotropicEoS}

A key step of any fluid model involves the connection to the microphysics as represented by the equation of state. As a first stab at this, let us outline the logic for the simple barotropic case (and then return to the issue in \cref{sec:2parEoS}, with a more realistic model in mind). In the case of a barotropic model the starting point is a one-parameter energy density $\varepsilon = \varepsilon(n)$, which leads to the thermodynamical (Gibbs) relation
\begin{equation}\label{eq:1parGibbs}
    p + \varepsilon = n\mu \;,
\end{equation}
where the chemical potential is defined as
\begin{equation}
    \mu = \frac{d \varepsilon}{dn} \;.
\end{equation}
Introducing fluctuations in all the scalars, as before (eg. $p=\overline p + \delta p$), we have
\begin{equation}
    \overline p + \delta p + \overline \varepsilon + \delta \varepsilon = (\overline n + \delta n ) (\overline \mu + \delta \mu) 
= \overline n\ \overline \mu + \overline n \delta \mu + \overline \mu \delta n + \delta n \delta \mu \;.
\end{equation}
Since the average of linear fluctuations are all taken to vanish in the present case, this leads to 
\begin{equation}\label{pgibbs1}
 \overline p  + \overline \varepsilon  
= \overline n\ \overline \mu + \overline{ \delta n \delta \mu }  \;, 
\end{equation}
which shows that the number density fluctuations impact on the equation of state inversion required to evolve the system. In order to proceed, we need to provide a closure relation for $\overline{ \delta n \delta \mu}$.

Noting that the energy was assumed to depend only on the number density, the  fluctuations in \eqref{pgibbs1} should not be independent. In order to take a closer look at this, we may Taylor expand for small fluctuations $\delta n$ (recalling that we have already considered expanding the Lorentz factor in this way). This then leads to 
\begin{equation}
\mu \approx \mu(\overline n) + \mu'(\overline n) \delta n + \frac{1}{2} \mu''(\overline n ) \delta n\delta n \;.
\end{equation}
That is, we identify
\begin{equation}
    \overline \mu =  \mu(\overline n)+  \frac{1}{2} \mu''(\overline n ) \overline {\delta n\delta n} \;,
\end{equation}
and
\begin{equation}
\delta \mu = \mu'(\overline n) \delta n \;. 
\end{equation}
The averaged Gibbs relation then becomes
\begin{equation}
    \overline p + \overline \varepsilon = \overline n \mu (\overline n) + \left[ \frac{\overline n} {2} \mu''(\overline n) + \mu'(\overline n)  \right] \overline { \delta n\delta n} \;, 
    \label{gibbs}
\end{equation}
showing that---in addition to the derivatives of the chemical potential--- we need to provide a (closure) relation for $\overline {\delta n\delta n}$. The other fluctuations are similarly slaved to $\delta n$. We get 
\begin{equation}
    \varepsilon(n) \approx \varepsilon(\overline n) + \varepsilon'(\overline n) \delta n + \frac{1}{2}\varepsilon''(\overline n) \delta n\delta n \;,
\end{equation}
which leads to
\begin{equation}
    \overline \varepsilon = \varepsilon (\overline n) + \frac{1}{2}\varepsilon''(\overline n) \overline{ \delta n\delta n }\;,
\end{equation}
and 
\begin{equation}
    \delta \varepsilon = \varepsilon'(\overline n) \delta n = \mu(\overline n) \delta n\;.
\end{equation}
 
In summary, we  have  equations describing the (proper time) evolution of  $\tilde n$, $\tilde \varepsilon$ and $\tilde u^a$, represented by equations \eqref{bary2}, \eqref{enav} and \eqref{momav}. In order to  solve these equations, we need to provide a closure relation via an equation of state, plus closure relations for the fluctuations involved in $\overline \gamma$, $q^a$ and $s^{ab}$, that is $\overline{\delta n \delta v^a}$ and $\overline{\delta v^a \delta v^b}$. Once these, and $\overline{\delta n\delta n}$, are provided, we can work out $\overline{n}$ from $\tn$ and---using the equation of state to give the derivatives of the chemical potential---then we have $\overline \varepsilon$, as well. Using either \eqref{gibbs} or \eqref{newp} we can then rewrite $\overline p$ in terms of resolved variables and the closure terms, which completes the set of quantities we need to close the system and carry on the evolution. In principle, the fibration based averaging model is complete.

\section{Fluid element filtering}
\label{sec:filter}

Having outlined a workable procedure for averaging in the spacetime setting, let us turn to the issue of filtering. While doing so, we will also take another step towards a practical implementation. Up to this point we have, mainly for pedagogical reasons, assumed that the averaging procedure is explicitly carried out (or at least that it can be). This was useful as it helped explain the underlying principles and how the relevant residuals arise. However, in an actual implementation it is  natural to view the residuals as additional quantities to be modelled in order to capture  the unresolved features.  This brings the required closure relations into focus.

Another reason for turning our attention to the filtering problem is that  most actual numerical simulations adopt this strategy rather than averaging. Formally, we expect the two problems to be similar, but we know from \cref{sec:AverVsFilter} that the filtering problem involves a slightly different logic. In particular, it is less natural---and also not desirable since the filtered fluctuations are unlikely to vanish---to consider an expansion in terms of the fluctuations. This is an important difference, so we need to carefully consider the steps in the analysis.

The natural way to proceed is to use a weighted average, in the spirit of \eqref{FavreU}. We would then start from\footnote{It is important to note that quantities like $\tilde n$ are not the same in the averaging and filtering cases. Still, we are using the same notation because they play the same role in each evolution scheme. Note also that our definition does not mean that $\la \tilde{u}^a \ra = \tilde{u}^a$. }
\begin{equation}
\langle n^a \rangle \equiv \tilde n \tilde u^a \ , 
\end{equation}
with (by construction) $\tilde u_a \tilde u^a=-1$,
leading to 
\begin{equation}
    \tilde n = - \tilde u_a \langle n^a \rangle \ .
\end{equation}
As in the averaging case, it is easy to see that the continuity equation then becomes
\begin{equation}
\la \nabla_a n^a\ra = \nabla_a \langle n^a \rangle = 0 \quad \Longrightarrow \quad \dot\tn+ \tilde n \nabla_a \tilde u^a = 0 \ .
\label{bary3}
\end{equation}

The main lesson is that, formally, the equation we arrive at takes the same form as in the averaging case; equation \eqref{bary2}. Still, there are differences relating to i) the  nonlinear quantities that have to be provided by a closure model and ii) the interpretation of the resolved/evolved variables. 

Deferring the discussion of more general aspects of  the equation of state for a moment (see \cref{sec:2parEoS}), let us move on to write down the equations of motion consistent with \eqref{bary3}. The filtered version of the perfect fluid stress-energy-momentum tensor can be written as
\begin{equation}
\langle T^{ab} \rangle 
= (\langle p \rangle + \langle \varepsilon \rangle ) \tilde u^a \tilde u^b + \langle  p \rangle g^{ab} + \tau^{ab} \ , 
\label{filtab}
\end{equation}
where  
\begin{equation}
\tau^{ab} =  \langle ( p + \varepsilon)  u^a  u^b \rangle - ( \langle p \rangle +  \langle \varepsilon \rangle ) \tilde u^a \tilde u^b 
\end{equation}
requires a closure relation.

Introducing, as before, the energy density measured by $\tilde u ^a$:
\begin{equation}
    \tilde \varepsilon = \tilde u_a \tilde u_b \langle T^{ab} \rangle = \langle \varepsilon\rangle + \tilde u_a \tilde u_b \tau^{ab} \ , 
\end{equation}
we can rewrite the filtered stress-energy-momentum tensor as
\begin{equation}
\label{eq:filtered_tab}
\langle T^{ab} \rangle 
= (\langle p \rangle + \tilde \varepsilon  ) \tilde u^a \tilde u^b + \langle  p \rangle g^{ab} + 2 \tilde u^{(a} q^{b)} + s^{ab} \ , 
\end{equation}
with
\begin{equation}\label{eq:filtered_tab_q}
    q^a = - \tilde \perp^a_b \tilde u_c \tau^{cb} = - \tilde \perp^a_b \tilde u_c \,\langle ( p + \varepsilon)  u^c  u^b \rangle\ , 
\end{equation}
and
\begin{equation}\label{eq:filtered_tab_sab}
    s^{ab} = \tilde \perp^a_c \tilde \perp^b_d \tau^{cd} = \tilde \perp^a_c \tilde \perp^b_d \, \langle ( p + \varepsilon)  u^c  u^d \rangle \ .
\end{equation}
It is easy to see that the energy and momentum equations  take exactly the same form (once we change $\langle p\rangle \to \overline p$) as in the averaged case (cf. equations \eqref{enav} and \eqref{momav}). As in  that case, the equations of motion provide all the information we need to solve the system once the relevant closure relations are provided. The key to a workable model is to provide appropriate closure relations, so it is important to understand what this involves. 

\section{Filtered Thermodynamics}
\label{sec:2parEoS}

So far we have only outlined the argument for the simple case of a barotropic fluid. This is not a realistic assumption for numerical simulations of nonlinear neutron-star dynamics. As a finite temperature description with a realistic matter composition is a prerequisite for relativistic models,  let us expand our model in this direction. This immediately leads to a conceptually important conclusion.
A quick look back at \cref{pgibbs1} shows that, even if we start from a barotropic Gibbs relation at the fine scale, the averaged/filtered result is effectively ``non-barotropic''. In fact, the  $\overline{\delta n \delta\mu}$ closure term could be interpreted as an 
``entropy-like'' contribution associated with the fluctuations. This suggests that it makes sense to start straight away from a non-barotropic model, i.e. with a Gibbs relation of form 
\begin{equation}\label{eq:2parGibbs}
	p + \veps = \mu n + T s \;,
\end{equation}
where $s$ is the entropy density and $T$ is the associated temperature.

The barotropic example highlights that the ``effective'' term that stems from the averaging or filtering procedure does not relate to the actual entropy, in the sense that it is not associated with some dissipative process and/or entropy production. This is evident from the fact that we have freedom in the choice of the averaging/filtering observer, and therefore in the variables to be evolved, and one might choose to frame the model in such a way that the fluctuation terms are reabsorbed in the definition of the variables themselves, just like we did for the weighted four-velocity $\tilde u^a$ in \cref{FavreU}.

Let us make these points more concrete by considering a model that is non-barotropic from the get-go. Because there is no formal difference in the resulting equations, we will do this without distinguishing between the averaging and filtering cases (although, as we are now familiar with the logic, using the slightly more abstract notation from the latter). 

We start by discussing some subtleties of the barotropic example that we previously left aside. The energy density of a barotropic fluid is a function of the matter density only, which means there is no need to evolve both quantities separately. On the fine-scale, the energy equation contains the same information as the continuity equation:
\begin{equation}
\begin{split}
	\frac{d\veps}{d\tau} + (p + \veps)\nabla_au^a = \mu \frac{d n}{d\tau} + \mu n\nabla_au^a = \mu (\nabla_an^a) = 0 \;,
\end{split}
\end{equation}
where $\tau$ is the proper time associated with the actual fluid worldlines with tangent $u^a$. Note that the Gibbs relation \cref{eq:1parGibbs} is crucial for this argument. However, the situation changes on the coarse-grained scale. The link between the micro-scale equation of state and the resolved energy is not trivial---it may have to be established by a set of high resolution simulations, even though this may not be practical/feasible. In effect, the equation of state we are working with at the filtered scale is not the one you get from nuclear physics.  This is, in fact, true also in the simpler case where we set to zero the contribution coming from the $\tau^{ab}$ residual as the evolved density is not obtained by averaging the fine-scale one, so $\tn \ne \la n \ra$. The net result is that we have to treat the resolved energy $\teps$ and the resolved density $\tn$ as independent variables, and evolve both of them. 

\subsection{The effective entropy}

The subtle difference in the counting of independent variables between the fine- and coarse scale models obviously no longer exists for a non-barotropic fluid. For a two-parameter equation of state, the energy and particle density can be taken as independent variables already at the fine scale. This is, indeed, standard practice in numerical relativity simulations.
Let us also recall that, if the fluid is ideal there is no additional information gained from evolving the entropy current. 
In fact, the (ideal-fluid) energy and continuity equations are
\begin{subequations}
\begin{align}
	\nabla_a n^a = \frac{d n}{d\tau}  + n \nabla_au^a &= 0 \;, \\
    \frac{d\veps}{d\tau}  + (p+\veps)\nabla_au^a &= 0
\end{align}	
\end{subequations}
while the entropy---now considered as a function of $\veps $ and $n$, $s=s(n,\varepsilon)$---is automatically advected at the fine scale level. That is, we have
\begin{equation} \label{eq:AdvectEntropyFine}
\begin{split}
T\nabla_a(su^a) &= T \frac{d s}{d\tau} + Ts\nabla_au^a = \bigg(\frac{d\veps}{d\tau} - \mu \frac{d n}{d\tau}\bigg) + (\veps + p - \mu n)\nabla_au^a \\
&= \left[\frac{d \veps}{d\tau} + (p+ \veps)\nabla_au^a\right] -\mu \left(\frac{d n}{d\tau} + n\nabla_au^a\right) = 0 \;.
\end{split}
\end{equation}

To complete the model set-up, we still have to clarify how the filtered pressure relates to the evolved variables. That is, we need some representation of the equation of state on the filtered scale. One of the main lessons from the  barotropic model is that the filtered Gibbs relation effectively takes the form \eqref{eq:2parGibbs}, even though the microphysics builds on a single-parameter model. One way to interpret this is that the filtering introduces ``non-ideal'' features that can be associated with an ``effective entropy''. To make progress in the general case, we may then---pragmatically---introduce thermodynamical relations at the level of  the coarse scale. The natural way to do this is to assume the same equation of state at the coarse scale, and let the residuals account for the additional contributions. As we are evolving $\tilde n$ and $\tilde \varepsilon$ we can work with a resolved entropy defined as the usual thermodynamical potential $\ts \doteq s(\teps,\,\tn)$. The resolved temperature and chemical potential then follow from the standard definitions
\begin{subequations}
\begin{align}
	\frac{1}{\tT} &\doteq \Big(\frac{\partial \ts}{\partial \tilde \veps}\Big)_{\tilde n}(\teps,\,\tn) \;,\\
    -\frac{\tmu}{\tT} &\doteq \Big(\frac{\partial \ts}{\partial \tilde n}\Big)_{\tilde \veps}(\teps,\,\tn) \;.
\end{align}
\end{subequations}
We stress that we choose to use the entropy as thermodynamic potential here, as it is a function of the chosen independent variables in the equations of motion, $\tilde n$ and $\tilde \veps$. Note also, that $\ts$ need not represent the true entropy and $\tT$ need not be the actual temperature, either. We are simply assuming that the usual thermodynamical definitions ``make sense'' at the filtering scale. This decision obviously has an impact on the interpretation of the results. This is a thorny issue which we will have to return to at a later point.

Following the same logic we used for the barotropic case (see \cref{subsec:BarotropicEoS}) we filter \cref{eq:2parGibbs} and rewrite it as
\begin{equation}\label{eq:FilteredPressure}
	\la p \ra= -\teps + \tmu\tn + \tT\ts  + M \;,
\end{equation}
with 
\begin{align}\label{eq:M_Tau_residuals}
    M &= \Big(\la Ts\ra - \tT\ts\Big) +\Big ( \la\mu n\ra  - \tmu\tn\Big) -\Big(\la\veps \ra - \teps\Big)\;.
\end{align}
The argument is now complete. We have explained how we may express the averaged/filtered pressure that enters the equations of motion in terms of the resolved variables $\teps,\,\tn$ and the (new) residual $M$. It is easy to see that the  results are similar to those for a non-ideal fluid (with heat flow and viscosity). In particular, in this comparison $M$ is linked to the bulk viscous pressure. 

Having considered the resolved thermodynamics, we can turn to the ``entropy production'' associated with the averaging/filtering procedure. The final evolution equations  \cref{enav,momav} clearly remind us of the result for a dissipative fluid (see \cite{livrev}) so we are motivated to consider possible constraints stemming from the second law of thermodynamics. To do this we can work through steps analogous to \cref{eq:AdvectEntropyFine}  to establish the impact of the averaging/filtering procedure. Because the resolved entropy $\ts$ is taken to be a function of the resolved energy $\teps$ and the number density $\tn$ we have 
\begin{equation}
    \tT \nabla_a (\ts\tu^a) = \tT \ts \nabla_a \tu^a + \tT \dot \ts = \tT \ts \nabla_a \tu^a + \dot\teps - \tmu\dot\tn \;.
\end{equation}
Now, by means of the filtered continuity and energy \cref{bary2,enav} we obtain 
\begin{equation}\label{eq:effective_s_production}
\begin{split}
     \tT \nabla_a (\ts\tu^a) &= \big( \tT\ts +\tmu\tn - \la p \ra - \teps \big)\nabla_a\tu^a - q^a\tilde a_a - \nabla_aq^a - s^{ab}\nabla_a\tu_b \\
     &= - M \nabla_a\tu^a - q^a\tilde a_a - \nabla_aq^a - s^{ab}\nabla_a\tu_b \;.
\end{split}
\end{equation}
This shows that the entropy is no longer advected at the coarse scale, as a result of the averaging/filtering procedure. However, the fine scale (``exact'') theory is ideal, so the actual entropy is advected (see \cref{eq:AdvectEntropyFine}). 
For this reason, the model is not constrained by the second law at the coarse-grained scale. This is a very important point as it impacts on the closure relations (see below), which (evidently) can be discussed without considering the thermodynamical restrictions for ``real'' dissipative fluids. Effectively, the heat-flux term in \cref{eq:average_q} or \cref{eq:filtered_tab_q} is associated with energy transfer from  large eddies to small ones (or vice versa) rather than being a faithful heat transfer. We also note that this would not change even if we started from a fluid that is dissipative already at the fine scale.  Restrictions stemming from the second law of thermodynamics apply only at the fine-scale level, not at the coarse one.

\subsection{Energy cascade argument}\label{sec:energy_cascade}

Having discussed the thermodynamical interpretation of the quantities that enter the equations of motion, it makes sense to consider the involved energy cascade. This is relevant because an analogous argument is used in standard work on turbulence (see, for instance, \cite{Lilly,Leonard}) to motivate the closure of the fluid equations. It is useful to spell out the relativistic analogue of the classical argument. 

The starting point is the energy equation \eqref{enav}) rewritten as 
\begin{equation}
    \underbrace{\dot\teps + (\la p\ra + \teps)\nabla_a \tu^a}_{\text{macro}} = \overbrace{- q^b \tilde a_b + \nabla_a q^a + \tu_b\nabla_a s^{ab}}^{\text{mixed}} \;.
\end{equation}
Here, we have highlighted that the terms on the left-hand side can be considered as macroscopic, in the sense that they involve only resolved quantities and describe an ideal evolution---intended to correctly capture the large-scale dynamics. In contrast, the terms on the right-hand side are ``mixed'' as they involve unresolved quantities---the residuals---and couple macro- and micro-scale terms. In effect, they can be thought of as transferring energy from one scale to another. 

To see this, we may, for a moment, assume a steady state evolution. As a consequence of the matter continuity equation, we then have $\nabla_a\tu^a = 0$ and therefore rewrite the energy equation as (setting to zero terms involving time derivatives with respect to $\tu^a$) 
\begin{equation}\label{eq:ClassicaCascade}
    \nabla_a q^a = s^{ab} \nabla_a\tu_b \;.
\end{equation}
In this relation, the term on the left-hand side should represent the energy sink (source) due to the (inverse) energy cascade---subtracting energy from the macro-scale into the micro one (or vice versa).
As an intuitive model, which ensures that the right-hand side of \eqref{eq:ClassicaCascade} does not become negative and draws on Newton's law of viscosity, Boussinesq suggested that one should relate the turbulent stress to the mean shear flow (see, e.g., \cite{mcdonough}). In our case, this leads  to
\begin{equation}\label{eq:sab_propto_sigma}
    s^{ab} \propto \tsig^{ab} \;,
\end{equation}
where 
\begin{equation}
    \tsig_{ab} = \left[  \tilde{\perp}^c_{(a} \tilde{\perp}^d_{b)}  - \frac{1}{3} \tilde{\perp}^{cd} \tilde{\perp}_{ab} \right] \nabla_c \tilde{u}_d.
\end{equation}
Motivated by this argument we move on to develop a closure scheme to complete the ``fibration framework'' we are proposing.

\section{An explicit closure model}\label{sec:closure}

Our ultimate aim is to develop a consistent scheme for large-eddy simulations in relativity. Even though this involves numerical aspects which we will not touch upon here (they are discussed in a follow-up paper \cite{fibrLESnum}), we need to provide a strategy for closing the system of equations already at the fibration level. This is the problem we focus on now.

As we have seen, in order to carry out an evolution we need to provide some prescription for the residual terms, that is $\tau^{ab},\,M$ in the filtering case or, equivalently,  $\overline{\delta n\delta n},\,\overline{\delta n\delta v^a},\,\overline{\delta v^a\delta v^b}$ in the averaging problem. In classical computational fluid dynamics, one of the earliest closures proposed---still widely used---is due to Smagorinsky \cite{Smagorinksy}. This model---which is motivated by the Boussinesq argument from the previous section---effectively boils down to retaining only the $s^{ab}$ term and modelling it as a traceless tensor proportional to the (resolved) shear-flow. 

We propose to model the residuals in terms of a general expansion in derivatives of the resolved variables, $\tn,\,\tu^a$ and $\teps$. For  practical reasons we halt the derivative expansion at first order\footnote{One might naively expect this to lead to issues relating to stability and causality, and that a closure scheme reminiscent of the second-order Israel-Stewart theory \cite{IsraelStewart79,IsraelStewart79bis} would be required to avoid such issues. However, as we will demonstrate, this is not the case.}, and decompose the gradients of the resolved quantities as
\begin{subequations}
\begin{align}
	\nabla_a \tn &= \tilde{\perp} ^b_a \nabla_b\tn - \tu_a \dot{\tn} \;, \\
    \nabla_a \teps &= \tilde{\perp} ^b_a \nabla_b\teps - \tu_a \dot{\teps} \;,\\
    \nabla_a\tu_b &= - \ta_b\tu_a + \tvort_{ab} + \tsig_{ab} + \frac{1}{3}\ttheta\tilde{\perp}_{ab} \;,
\end{align}
\end{subequations}
where, as before, $\dot{\tilde n} = \tilde u^a\nabla_a \tilde n$ (similarly for $\dot{\tilde \veps}$) and
\begin{subequations}
\begin{align}
    \ta_a &= \tu^b\nabla_b\tu_a \\
    \tsig_{ab} &=\tperp^c_{(a} \tperp^d_{b)} \nabla_c \tu_d -\frac{1}{3}\ttheta \tperp_{ab} \\
    \tvort_{ab} &= \tperp^c_{[a} \tperp^d_{b]} \nabla_c \tu_d \; , \\
    \ttheta &= \tperp^{ab}\nabla_a \tu_b\; .
\end{align}
\end{subequations}
This closure scheme is analogous, although in a different spirit, to the most general constitutive relations discussed for dissipative hydrodynamics (at the linear level), see \cite{KovtunStable,HoultKovtun2020}. Let us also stress that while the closure scheme we are  proposing is built from gradients of the (resolved) fluid variables, it is important not to confuse it with the model discussed in \cite{viganoNR,carrasco,viganoGR}. 

Because there is no formal difference in the modelling of the sub-filter scale terms between the averaging and filtering cases, let us set up the closure scheme for the filtering case. We also immediately consider the case of a two-parameter equation of state, as the barotropic limit can be easily recovered from the more general results. We then have to model the residuals $q^{a} \text{ and }s^{ab}$. Recalling the definitions \cref{eq:filtered_tab_q,eq:filtered_tab_sab} we express these as
\begin{subequations}\label{eq:KovtunClosure_qs}
\begin{align}
    s^{ab} &= -\eta\tsig^{ab} + (\pi_1 \ttheta + \pi_2 \dot \tn + \pi_3 \dot \teps) \tilde{\perp}^{ab}\;,\\
    q^a &= \theta_1 \ta^a + \theta_2 \tilde{\perp}^a_b\nabla^b\tn + \theta_3\tilde{\perp}^a_b\nabla^b\teps\;.
\end{align}
\end{subequations}
In order to evolve the system, we also need to express $\la p\ra$ in terms of the resolved variables. To do so, we have to provide $M$. As this is a scalar, we model it as (see \cref{eq:M_Tau_residuals})
\begin{equation}\label{eq:KovtunClosure_Mtau}
	M = \chi_1 \ttheta + \chi_2 \dot \tn + \chi_3 \dot \teps \;.
\end{equation}
We have now introduced a total of $10$ parameters to be used in the actual large-eddy model. These parameters will generally be functions of the evolved variables, the form of which may possibly be inferred from high-resolution simulations,  but for simplicity we will take them to be constant in the following. 
 
\subsection{Stability Analysis}\label{subsec:stability}

Let us turn to the issue of linear stability, as this is a necessary condition for the system of equations to be (numerically) solved. Another motivation is the fact that the (resolved scale) ``effective'' theory we have arrived at resembles that of a dissipative fluid, and it is well known that the standard/textbook relativistic viscous hydrodynamics equations are unstable (see \cite{HiscockInsta}). 

The linear stability of the effective theory obviously depends on the closure used, so let us focus on the specific relations proposed above. However, the aim is not to discuss the stability of the closure model in full generality, only to provide a ``proof of principle'' argument that there are choices for the parameters for which the model is linearly stable. 

As the averaging/filtering residuals have been expressed in terms of gradients of the evolved variables and we are considering a local region, it makes sense to assume that the background configuration---the stability of which we want to study---is that of a homogeneous fluid at rest. We will also consider, as usual, a flat background spacetime and ignore metric perturbations. This is justified---even in the general relativistic context---since the stability analysis is intended to be local, so that the effects of gravity can be transformed away (using a local inertial frame argument, in the spirit of the Fermi frame logic). Finally, we simplify the notation in order not to clutter up the equations. We drop the ``tildes'' used to identify the resolved variables, as we no longer need to make the distinction. Instead, we identify background quantities with a subscript ($0$).  For instance, we write the background four-velocity as $u_0^a$ and the chemical potential in the background configuration as $\mu_0$.

Let us start by expanding the perturbed fields (indicated by a $\delta$) in Fourier modes:
\begin{subequations}
\begin{align}
    \delta u^a &= \B^a \pw \;,\\
    \delta n &= \A \pw\;,\\
    \delta \veps &= \E \pw \;.
\end{align}
\end{subequations}
where $\B^a$ is orthogonal to $u_0^a$ because $u^a_0\delta u_a = 0$ as a result of the four-velocity normalization. We also decompose the wave-vector $k^a$ as
\begin{equation}
     k^a = \omega u_0^a + k\hat k^a \;,
\end{equation}
where $\omega$ is the frequency, $k$ is the wavelength and $\hat k^a$ is a unit four-vector orthogonal to $u_0^a$ which describes each mode's direction. Because of the metric signature convention (+2), the system will be linearly stable (in time) if all solutions to the dispersion relation---written as $\omega = \omega(k)$---have a negative (or vanishing) imaginary part. We will also use $\hk^a\hk^b$ and $\delta^{ab} -\hk^a\hk^b$ to decompose the momentum equation as well as $\B^a= (0,\,\B_L,\,\B_{T1},\,\B_{T2})^\top$ into its longitudinal and transverse part (with respect to wave direction).

In order to write the linearized equations in terms of the perturbed fields, we have to clarify how to perturb the pressure. As can be seen from \cref{eq:FilteredPressure}, its explicit expression depends on $M$. Let us first focus on the non-residual contribution and come back to $M$ later:
\begin{equation}
    \delta p =  (\C \A +\D\E)\pw + \delta M\;,
\end{equation}
where we have defined 
\begin{subequations}\label{eq:CDdefinitions}
\begin{align}
    \C &= \Big(\frac{\partial p}{\partial n}\Big)_\veps (n_0,\,\veps_0) \;,\\
    \D &= \Big(\frac{\partial p}{\partial \veps}\Big)_n(n_0,\,\veps_0) 
\end{align}
\end{subequations}
to simplify the expressions that follow.

Let us now start by linearizing first the non-residual part of the equations of motion. The result is 
\begin{subequations}
\begin{align}
    &-i\omega \A + i k n_0 \B_L = 0\;,\\
    &-i\omega \E + i h_0 k \B_L = 0\;, \\
    &-i h_0 \omega\B_L + ik(\C \A +\D\E) = 0\;, \\
    &-i h_0 \omega\B_{T1} = 0\;,\\
     &-i h_0 \omega\B_{T2}  = 0\;,
\end{align}
\end{subequations}
where we have introduced the usual enthalpy density, $h_0 = p_0 + \veps_0$. To work out the full linearized system of equations, we consider each residual at a time. We start by looking at the trace-free part of $s^{ab}$, as this would correspond to the (fibration version of the) model  proposed by Smagorinsky in the Newtonian context. A straightforward calculation leads to
\begin{equation}
\begin{split}
    \delta\sigma^{ab} &= \delta\big(\perp^{a}_{c}\perp^{b}_{d}\partial^{(c}u^{d)} - \frac{1}{3} (\partial_cu^c)\perp^{ab}\big) \\
    &= ik\big(\hat{k}^{(a}\B^{b)} - \frac{1}{3}\B_L \perp_0^{ab}\big)\pw \;,
\end{split}
\end{equation}
where $\perp_0^{ab} = \eta^{ab}+ u_0^a u_0^b$ is the projection orthogonal to the background velocity and $\eta^{ab}$ is the Minkowski metric. As for the trace part of $s^{ab}$ we have  
\begin{equation}
\begin{split}
	&\delta\Big[\big(\pi_1 \partial_cu^c + \pi_2 u^c\partial_c n + \pi_3 u^c\partial_c\veps\big) \perp^{ab}\Big] = \\
    &= \big(i \pi_1 k \B_L - i\pi_2 \omega \A - i\pi_3 \omega\E\big) \perp_0^{ab} \;,
\end{split}
\end{equation}
and it is easy to see that these additional terms only affect the longitudinal projection of the momentum equation. Next we have the heat flux $q^a$. It is fairly easy to see that only two (out of five) terms will contribute to the linearized equations. These terms lead to 
\begin{equation}
	\delta (\partial_aq^a) = +\theta_1 \omega k \B_L - \theta_2 k^2 \A - \theta_3 k^2\E
\end{equation}
entering the energy equation, and 
\begin{equation}
	\delta (\perp^b_c u^a\partial_aq^c) = -\theta_1 \omega^2 B^b + \theta_2 \omega k\hk^b\A + \theta_3 \omega k\hk^b\E
\end{equation}
entering the momentum equation. We note that the last two terms in the expression above affect only the longitudinal projection, while the first term modifies both the longitudinal and transverse components. Last but not least, we consider the residuals that arise from the Gibbs relation \eqref{eq:KovtunClosure_Mtau}.
It is easy to see that this residual will not affect the (linearized) energy equation, while it contributes to the longitudinal momentum equation as
\begin{equation}
	\delta \big(\perp^{ab}\partial_a M\big) = \big[-\chi_1k^2\B_L + \chi_2 k\omega \A + \chi_3 k\omega \E\big] \hk^b\;.
\end{equation}
Collecting everything together, we can write the linearized equations\footnote{Note that the coefficient matrix depends only on $\omega = -u^ak_a$ and $k^2 = k^ak_a + \omega^2$, in accordance with Lorentz invariance.} as 
\begin{equation}\label{eq:GeneralMatrix}
    \begin{pmatrix}
    \pmb{L} & \pmb{0} \\
    \pmb{0} & \pmb{T} 
    \end{pmatrix}
    \,  \begin{pmatrix} 
    \A & \E &\B_L & \B_{T1} & \B_{T2} 
    \end{pmatrix}^\top = 0 \;,
\end{equation}
where
\begin{equation}\label{eq:GenLongMatrix}
    \pmb{L}  = \begin{pmatrix}
    -i\omega  & 0 & i n_0 k  \\
    -\theta_2 k^2  & -i\omega -\theta_3k^2 & i h_0 k +\theta_1 k\omega \\
     ik\C + (\zeta_2+ \theta_2) k\omega & i \D k + (\zeta_3 + \theta_3) k\omega& -\big(i h_0\omega -\frac{2}{3}\eta k^2 + \zeta_1 k^2 + \theta_1 \omega^2\big) \\
    \end{pmatrix}
\end{equation}
and 
\begin{equation}\label{eq:GenTransverseMatrix}
    \pmb{T}  = \begin{pmatrix}
    -\big( i h_0\omega - \frac{\eta}{2} k^2 + \theta_1\omega^2 \big) & 0 \\
    0 &-\big( i h_0 \omega - \frac{\eta}{2} k^2 + \theta_1\omega^2 \big)
    \end{pmatrix}
\end{equation}
and we have introduced $\zeta_i =  \chi_i  + \pi_i$ with $i = 1,2,3$. The proposed gradient expansion and the resulting equations are similar to those of \cite{KovtunStable,HoultKovtun2020}. However, the parameters of the two models are  not in one-to-one correspondence---mainly because the gradient expansion  is carried out in terms of different variables. Some level of care is required when comparing the results. 

Before we proceed, let us stress (again) that, because the entropy $\ts$ does not represent the true one, we are allowed to violate the second law of thermodynamics. This will give us more freedom (with respect to faithful dissipative fluids) to control the stability of the closure. 

\subsection{Smagorinsky model}\label{subsec:Smagorinsky}

The stability analysis for the general case is perhaps best considered numerically, but we can make sufficient progress to understand the issues involved from simplified models. 
As a first step, let us consider the simple case where the only non-vanishing parameter in \cref{eq:KovtunClosure_qs,eq:KovtunClosure_Mtau} is $\eta$. This would correspond to the (fibration version of the) model proposed by Smagorinsky in the Newtonian context. Starting from  \cref{eq:GeneralMatrix} we easily obtain the linearized equations for this case
\begin{equation}
    \begin{pmatrix}
    -i\omega & 0 & i n_0 k & 0 & 0 \\
    0 & -i\omega & i h_0 k &0 & 0   \\
    ik\C & i\D k& -i h_0 \omega + \frac{2}{3}\eta k^2 & 0 & 0 \\
    0 & 0 & 0 & -i h_0 \omega + \frac{\eta}{2} k^2 & 0  \\
   0 & 0 & 0 & 0  & -i h_0 \omega + \frac{\eta}{2} k^2   
    \end{pmatrix} \,
    \begin{pmatrix}
    \A \\ \E \\ \B_L \\ \B_{T1} \\ \B_{T2} 
    \end{pmatrix} = 0\;.
\end{equation}
The required dispersion relations are obtained by setting to zero the determinant of the coefficient matrix. Working this out, we find that the transverse modes decouple, and the corresponding dispersion relation is 
\begin{equation}\label{eq:SmagorinskyTransverse}
    \omega = - i \frac{\eta}{2 h_0} k^2 \;.
\end{equation}
The other non-trivial modes are longitudinal, with dispersion relation 
\begin{equation}
 h_0 \omega^2 + i\frac{2}{3} \eta k^2\omega  - (h_0 \D + \C n_0)k^2 = 0\;.
\end{equation}
Solving this equation we obtain 
\begin{equation}\label{eq:SmagorinskyLongitudinal}
	\omega = \pm c_s k - i \frac{\eta}{3 h_0}k^2 + \mathcal{O}(k^3) \;,
\end{equation}
where
\begin{equation}\label{eq:Lmode_speed}
    c_s^2 = (\D + \C n_0/h_0) 
\end{equation}
is the usual sound speed. The longitudinal mode represents sound waves, while the transverse modes are not propagating. Both sets of modes are stable for $\eta>0$.  As a simple consistency check, it is easy to see that in the ideal limit, when $\eta = 0 $, one obtains a single non-trivial solution representing an undamped sound wave. 

The result demonstrates that the simple Smagorinsky model is stable according to the fibration observer\footnote{The stability result also lends support to the large-eddy model of \cite{radice1}, which implements a simple Smagorinsky closure in the foliation frame, where the simulation is then performed.}. This is a key conclusion for cosmological applications (see \cite{EllisInhomCosmo}), as these tend to involve a cosmological time associated with a co-moving observer; in essence, a fibration. The situation is different for numerical relativity simulations, which tend to be based on a spacetime foliation. The matter description  (formally) involves a fibration associated with fluid element worldlines, but the evolution is carried out using a non-comoving background. Our stability demonstration does not (yet) cover this case. In order to complete the argument,  we have to consider the stability issue in a different frame. We therefore introduce (as usual) the Eulerian observer $N^a$ as 
\begin{equation}
    u^a = W\left(N^a + v^a\right) \;, \quad \mbox{with} \quad
    W = (1 - v^2)^{-1/2} \;,
\end{equation} 
and note that  the two frames are related by a  Lorentz boost. 
This turns out to cause trouble. The simple Smagorinsky model, while  stable in the fibration frame, becomes unstable in the boosted frame. 

In order to demonstrate this result, we start by noting that (using primes to indicate boosted quantities)
\begin{subequations}
\begin{align}
    \partial'_a\veps' &= \Lambda^b_a\partial_b\veps = 0 \\
    \partial'_a n' &= \Lambda^b_a\partial_b n = 0 \\
    \partial'_a u'_b &= \Lambda^c_a\Lambda^d_b\partial_cu_d = 0 
\end{align}
\end{subequations}
where $\Lambda$ is the Lorentz boost matrix. As we are linearizing with respect to a homogeneous (in spacetime) background this confirms that the gradient-based closure scheme we are proposing  still makes sense. We have to work out the dispersion relations in a non-comoving frame, but because these are expressed in terms of $\omega = -k^au_a$ and $k^2 = k^a k_a + \om^2$, we just have to boost these quantities. We can then take (without loss of generality) $v^a$ to be in the $x$-direction, while $\hat k$ lies in the $x-y$ plane. Then we introduce the angle $\phi$ between the wave-vector and $v^a$ as $\hat k^a v_a = v\,\cos{\phi}$ and write the Lorentz boost as:
\begin{subequations}\label{eq:LorentzBoost}
\begin{align}
    \om &= W(\om' - vk'\cos{\phi}) \;, \\
    k_x &= W(k'\cos{\phi} - v\om') \;,\\
    k_y &= k'\sin{\phi} = k'_y \;,\\
    k_z &= 0 = k'_z\;.
\end{align}
\end{subequations}

Applying this to the transverse dispersion relation in \cref{eq:SmagorinskyTransverse} we obtain (dropping the primes for clarity)
\be
     \left(\eta W^2 v^2\right)\om^2 -2 \left(i h_0W - \eta W^2 v k\cos{\phi}  \right)\om  \\ -\left( \eta W^2 k^2\cos^2{\phi}+ \eta k^2 \sin^2{\phi}- 2ih_0\cos{\phi} Wvk\right) = 0
\ee
We note that \cref{eq:SmagorinskyTransverse} was a first order polynomial, while the boost made it second order, thus generating an additional solution. For long wavelengths, the two solutions are
\begin{subequations}
\begin{align}
    \om &= vk\cos{\phi} - i \frac{\eta}{2h_0 W^3} \left(\cos^2\phi + W^{-2}\sin^2\phi\right) k^2 + \O (k^3)\\
    \om &= i\frac{2h_0}{\eta W v^2} + \O(k)\;.
\end{align}
\end{subequations}
The first solution is the boosted version of the mode we obtained in the fibration. It is stable for $\eta> 0$ (as in the comoving frame), propagating with phase velocity $v\cos{\phi}$ and the decay rate reduces to the original value as $v\to 0,\,W\to 1$. There is, however, an additional solution which is non-vanishing for $k=0$ (in \cite{KovtunStable,HoultKovtun2020,FrontiersGavassinoAntonelli} these are referred to as ``gapped'' modes). This second mode is evidently unstable for $\eta > 0$. 

This result demonstrates that the simple Smagorinsky model is unstable when ``observed'' from a non-comoving frame. This is a well-known problem of the classic Eckart-Landau models for dissipative fluids \cite{HiscockInsta}. For real dissipative systems, stability of equilibrium is not only required for numerical implementations, it is also linked to intrinsic consistency. A system slightly out of equilibrium must evolve, ``by definition'', towards thermodynamical equilibrium, no matter if the fluid in equilibrium is at rest or not. Our case is different. However, because we are setting up the filtering scheme in the fibration, while the simulations will be carried out in the foliation, we  have to ensure that the model is ``covariantly'' stable.

\subsection{Fixing the Smagorinsky instability}\label{subsec:fixSmag}

The aim now is to show how we can fix the instability problem (in the boosted frame) by introducing more parameters in the closure model. Focusing first on the transverse modes, we see from \cref{eq:GenTransverseMatrix} that the only way to fix the problem is by considering a non-zero $\theta_1$. The co-moving transverse dispersion relation then becomes 
\begin{equation}
    2ih_0 \om - \eta k^2 + \theta_1 \om ^2 = 0 \;,
\end{equation}
with solutions for long wavelengths (i.e. small $k$):
\begin{subequations}
\begin{align}
    \om_+ &= -i \frac{\eta}{2h_0}k^2 + \O(k^4) \;,\\
    \om_- &= -2 i \frac{h_0}{\theta_1} + \O(k^2)\;.
\end{align}
\end{subequations}
Because the dispersion relation is now quadratic we obtain two solutions: the ``un-gapped'' mode from  the Smagorinsky model, and an additional gapped mode that appears already in the co-moving frame. The (long wavelength) stability in the un-boosted frame  is guaranteed by taking $\eta>0 $ (as before) alongside $\theta_1>0$. We can further check the stability in the co-moving frame at all wavelengths by means of the Routh-Hurwitz criterion (see, for instance \cite{korn2013mathematical}). In order to do so we introduce $\Delta = - i \om$ (to deal with real algebraic equations) and rewrite the dispersion relation as 
\begin{equation}
   \theta_1 \Delta^2 + 2 h_0 \Delta + \eta k^2 = 0 \;.
\end{equation}
Stability requires the solutions to have negative real part $\text{Re} \Delta< 0$. The Routh-Hurwitz criterion then guarantees the stability (at all wavelengths) as long as $\theta_1 > 0$ and $\eta > 0$. These conditions are identical to the ones obtained at long wavelengths. 

In order to check the stability in the boosted frame, we boost the transverse modes dispersion relation (as before) to get
\begin{multline}\label{eq:transverseBoosted}
     \left(\theta_1-\eta v^2  \right)W^2\om^2 + 2 \left(i h_0W + (\eta + \theta_1) W^2 v k\cos{\phi}  \right)\om  \\
    - \left(\eta W^2 k^2\cos^2{\phi}  + \eta k^2\sin^2{\phi} + 2 i h_0 W v k - \theta_1 W^2 v^2 k^2\cos^2\phi\right) = 0 
\end{multline}
To work out the long wavelength stability conditions, we may solve this equation perturbatively. This means that we introduce
\begin{subequations}\label{eq:IterativeExpansion}
\begin{align}
    \om &= \om_0  + \om_1 k+ \om_2 k^2+ \om_3 k^3 + \O(k^4)\\
    \om^2 &= \om_0^2+ 2\om_0\om_1 k + (\om_1^2 + 2 \om_0\om_2 )k^2 + (2\om_1\om_2 + 2 \om_0\om_3) k^3 + \O(k^4) \\
    \om^3 &= \om_0^3+3\om_0^2\om_1k + (3\om_0\om_1^2 + 3 \om_0^2\om_2) k^2 + (\om_1^3 + 6\om_0\om_1\om_2 + 3 \om_0^2\om_3)k^3 + \O(k^4) 
\end{align}
\end{subequations}
and solve order by order. Solving \cref{eq:transverseBoosted} to lowest order we find two solutions: the first is  the un-gapped mode ($\om_0$ = 0) while the second is given by
\begin{equation}
    \om_0 = - i \frac{2h_0 W^{-1}}{\theta_1 - \eta v^2 }  \,.
\end{equation}
We may focus on the un-gapped mode as we already have the imaginary part (to lowest order) for the gapped mode. Working to first order we obtain the phase velocity, while at second order we get the damping rate. Collecting the results, the small $k$ solutions to the boosted dispersion relation can be written
\begin{subequations}
\begin{align}
    \om &= - i \frac{2h_0 W^{-1}}{\theta_1 -\eta v^2} +  \O(k) \\
    \om &= v k \cos\phi + i \frac{\eta}{2h_0W^3}\left(\cos^2\phi + W^{-2}\sin^2\phi\right) + \O(k^3)
\end{align}
\end{subequations}
We  see that stability in the boosted frame requires $\eta>0$ and $\theta_1> \eta v^2$. 
To the best of our knowledge, stability at all wavelengths cannot be studied analytically in the boosted case. This is because the Routh-Hurwitz criterion applies to real polynomials only. The stability is then perhaps  best studied numerically, once a specific equation of state has been chosen. The main point here is that the fibration-based large-eddy model can be made to pass the key stability tests choosing appropriate parameters.

Next, we  turn our attention to the longitudinal modes, assuming again that the only non-vanishing parameters in \cref{eq:GenLongMatrix} are $\eta$ and $\theta_1$. The (comoving) longitudinal modes dispersion relation is then 
\begin{equation}\label{eq:ComovingLDisp}
    \om \left[\theta_1 \om^3 + i h_0 \om^2 - \left(\frac{2\eta}{3} +\theta_1 \D\right)k^2 \om - i h_0 c_s^2 k^2 \right] = 0\;.
\end{equation}
As before, we can work out the non-trivial longitudinal modes  using \cref{eq:IterativeExpansion}. Again, working to lowest order we find that one mode is ``gapped'' and two are not. These long wavelength modes are 
\begin{subequations}
\begin{align}
    \om &= -i \frac{h_0}{\theta_1} + \O(k^2) \;, \\
    \om &= \pm c_s k - \frac{i}{3h_0} \left(\eta - \frac{3}{2}(c_s^2 - \D)\theta_1 \right)k^2 + \O(k^3) \;.
\end{align}
\end{subequations}
In order to make sure the longitudinal modes are also stable we have to take $\theta_1 >0,\, \eta > 0$ and 
\begin{equation}\label{eq:UnboostedStabL+T}
    \frac{3}{2} (c_s^2 - \D) < \frac{\eta}{\theta_1} < \frac{1}{v^2} \;.
\end{equation}
We can then study the stability condition at all wavelengths using the Routh-Hurwitz criterion. To do so, we again introduce $\Delta = -i\om$ in \cref{eq:ComovingLDisp} to make it a real algebraic equation: 
\begin{equation}
    \theta_1 \Delta^3 + h_0 \Delta^2 + A \Delta k^2 + h_0 c_s^2 k^2 = 0 \;,
\end{equation}
where $A = 2/3\eta + \theta_1 \D$. From this it is easy to see that the Routh-Hurwitz criterion guarantees stability for \cref{eq:UnboostedStabL+T}. Again, the general $k$ case does not change the stability requirements. 

\begin{figure}
\centering
\includegraphics[width=\textwidth]{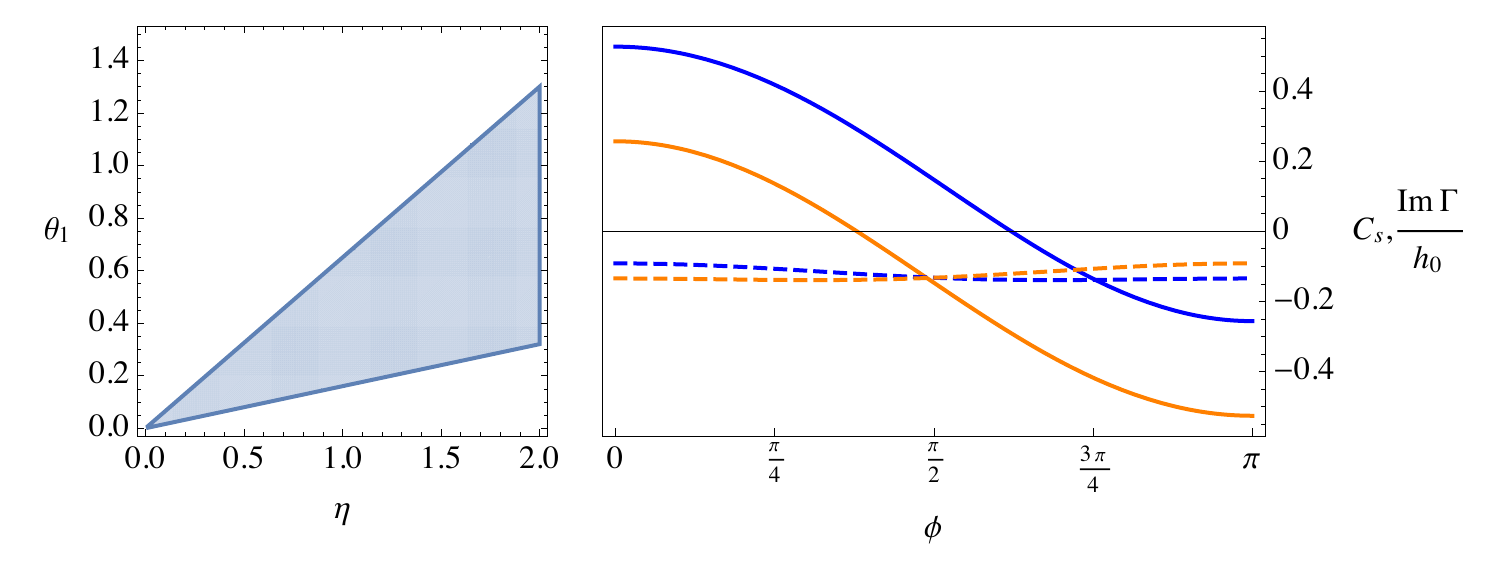}
\caption{
Left: shaded region resulting from combining  the stability constraints obtained for the  i) transverse modes (gapped, un-gapped, boosted, un-boosted), ii) longitudinal un-boosted modes (gapped and un-gapped), and iii) longitudinal boosted gapped modes. 
Right: sound speeds (solid) and damping rates (dashed) for the un-gapped longitudinal modes in the boosted frame. Colours match sound speeds with the corresponding damping rates. 
For illustrative purposes we used: $c_s = 0.16,\, v= 0.4,\,\eta = 1.2,\,\theta_1 = 0.5$ and a barotropic equation of state (both figures). 
}
\label{fig:BothPlots}
\end{figure}

As in the case of transverse modes, the story does not end here. We still have to establish the stability in the boosted frame. To do so, we ``boost'' \cref{eq:ComovingLDisp} using \cref{eq:LorentzBoost} to obtain 
\begin{equation}
    a\,\om^3 + b\, \om^2 + c\,\om + d = 0 \;,
\end{equation}
where
\begin{subequations}
\begin{align}
    a &= W^3 \left(\theta_1 - A\,v^2\right)  \;,\\
    b &= - W^2 \left[(3\theta_1 - 2A) Wvk\cos\phi - ih_0 - WAv^3k\cos\phi + ih_0c_s^2v^2\right] \;,\\
    c &=\begin{multlined}[t] 
    W \Big[(3\theta_1 - 2A)W^2v^2k^2\cos^2\phi - Ak^2\left(1 + W^2v^2 \cos^2\phi\right) \\- 2ih_0 Wv k (1 - c_s^2)\cos\phi\Big]
    \end{multlined}  \\
    d &=\begin{multlined}[t]
        - \Big[\theta_1 v^3 W^3\cos^3\phi -W A v k^3 \cos^3\phi - W^3 v^3 A k^3 \cos^3\phi \\
    -i h_0k^2 \left(W^2(v^2 - c_s^2)\cos^2\phi - c_s^2(1 - \cos^2\phi)\right)\Big] .
    \end{multlined}
\end{align}
\end{subequations}
Again we solve the problem using the expansion in \cref{eq:IterativeExpansion}. At lowest order we find two ``un-gapped'' modes and one ``gapped'' solution. The latter is given by 
\begin{equation}
    \om = - i \frac{h_0(1-c_s^2v^2)}{(\theta_1 - Av^2)W} + \O(k)
\end{equation}
which is stable for 
\begin{equation}\label{eq:BoostedGappedLStab}
    \eta < \frac{3}{2}\theta_1 \frac{1- \D v^2}{v^2} \;.
\end{equation}
As in the case of the ``un-gapped'' modes, working to $\O(k^2)$ (as the $\O(k)$ problem is trivial) we obtain the boosted sound speed $\om_1 = \C_s$. This is found by solving 
\begin{equation}
    W^2 (1- v^2c_s^2)\C_s^2 - 2W^2 v \cos\phi(1- c_s^2) \C_s + W^2( v^2 - c_s^2)\cos^2\phi  - c_s^2 \sin^2\phi = 0\;.
\end{equation}
To see that this result actually makes sense, we provide the solution for the two cases where $v^a$ is parallel/orthogonal to $\hat k^a$ (respectively)
\begin{subequations}
\begin{align}
    \C_s &= \frac{v\pm c_s}{1\pm v\,c_s} , \\
    \C_s &= \pm\frac{\sqrt{1- v^2}}{\sqrt{1- v^2c_s^2}}c_s .
\end{align}
\end{subequations}
The solution for different values of $\phi$  is best understood by considering specific examples, see \cref{fig:BothPlots}, noting that it only depends on the thermodynamic speed of sound $c_s$ and the relative velocity $v$. 
 
In order to work out the longitudinal mode damping we work at $\O(k^3)$. This  leads to a purely imaginary $\om_2 = \Gamma(\phi)$. The solution involves the boosted sound speed $\C_s$, and it is not particularly illuminating, so  it is also best understood by specific examples, see \cref{fig:BothPlots}. 

As we  see from the illustrations in \cref{fig:BothPlots} there are regions of the ($\eta,\theta$) parameter space  where all  modes are stable. As we are only aiming at a  proof of principle (not a comprehensive stability analysis) this concludes the argument. The stability conditions depend on the equation of state (which enters through $c_s,\,\D,\,\C$) and the relative velocity $v$. Therefore, an exhaustive study of the stability in the  general case is  best done once a specific equation of state has been chosen, following the logic outlined here. 

\section{Remarks and Conclusions}\label{sec:remarks}

Averaging and filtering are the standard strategies for dealing with the problem of simulating (computationally demanding) turbulent flows. Both approaches are complicated by the covariance of general relativity, where the split between space and time is an observer-dependent notion. As the problem is beginning to be considered from the point of view of numerical relativity \cite{Giacomazzo,radice1,radice2,aguil,viganoGR, duez}, it is important to understand the underpinning theory \cite{eyink}. Hence, we decided to start from the beginning and  considered how the different strategies should be implemented in the curved spacetime setting  relevant for, say, binary neutron-star mergers. 

After clarifying the sense in which consistency with the principles of general relativity poses interesting foundational questions, we argued that it is natural to set up the analysis in the ``fibration'' associated with individual fluid elements.
This then allowed us to introduce a meaningful local analysis via the use of Fermi coordinates \cite{fermi1,fermi2}, which defines the covariant averaging/filtering procedure. Building on this, we worked out the coarse-grained fluid dynamics, and considered  the impact  averaging/filtering has on the (thermodynamical) interpretation of the resolved variables.
Finally, because smoothing the fluid dynamics inevitably introduces a closure issue, we proposed a closure scheme and discussed its linear stability. This  completed the formal development of the fibration-based model. 

In order for this work to have practical relevance, we need to make contact with actual  simulations. This (typically) requires the introduction of a foliation and the associated ``3+1'' space-time split \cite{Gourgoulhon3+1, RezzollaZanotti}. 
This adds a new (Eulerian) observer to the game and brings new aspects to the discussion. In principle,  one might make progress simply by ``translating'' our present results to the foliation picture, but it is by no means clear that this is a sensible way to proceed. Moreover, any discussion of numerical simulations should consider a number of additional issues, like the role of numerical discretisation errors. As these are important aspects that need to be considered carefully, and it is important not to confuse the formal points we have discussed here with practical implementation questions, we  explore them in a separate  study \cite{fibrLESnum}. The work we have done so far takes us several steps towards the final destination, but we still have some way to go.

\acknowledgments

NA and IH acknowledge support from STFC via grants ST/R00045X/1 and ST/V000551/1.

\bibliography{averaging.bib}

\end{document}